\documentclass[runningheads]{llncs}
\newcommand{\etal}{\textit{et al.}}
\usepackage{url}
\usepackage{graphicx}
\usepackage[normalem]{ulem}
\usepackage{tabularx}
\usepackage{xcolor}
\usepackage{listings}
\usepackage{hyperref}

\hypersetup{
    colorlinks=true,
    linkcolor=black,  % Links within the document
    urlcolor=black,   % URLs
    citecolor=black,  % Citations (if using the `cite` package)
    pdfborder={0 0 0}, % Remove all borders
}

\definecolor{codegreen}{rgb}{0,0.6,0}
\definecolor{codegray}{rgb}{0.5,0.5,0.5}
\definecolor{codepurple}{rgb}{0.58,0,0.82}
\definecolor{backcolour}{rgb}{0.95,0.95,0.92}
\lstdefinestyle{mystyle}{
    commentstyle=\color{codegray},
    keywordstyle=\color{blue},
    numberstyle=\footnotesize,
    stringstyle=\color{codegreen},
    basicstyle=\ttfamily\footnotesize,
    breakatwhitespace=false,
    breaklines=true,
    captionpos=t, % bottom (b) or top (t)
    keepspaces=true,
    numbers=left,
    numbersep=5pt,
    showspaces=false,
    showstringspaces=false,
    showtabs=false,
    tabsize=2,
    xleftmargin=5.0ex,
    linewidth=\linewidth,
    language=python,
    escapeinside={(*@}{@*)}
}

\usepackage{booktabs}

\usepackage{etoolbox}
\AtBeginEnvironment{quote}{\itshape}

\usepackage{color}

\newcommand{\circled}[1]{\raisebox{.5pt}{\textcircled{\raisebox{-.9pt} {#1}}}}

\newcommand{\rqone}{\textbf{RQ1. Can generative AI be used to automate TDD? }}
\newcommand{\rqtwo}{\textbf{RQ2. What kind of interaction model between generative AI and human developers is more promising?}}

\usepackage{makecell}

\usepackage{amsmath}
\begin{document}

\makeatletter
\renewcommand\thelstlisting{\@arabic\c@lstlisting}
\makeatother
\mainmatter              % start of a contribution
\title{Generative AI for Test Driven Development: Preliminary Results}
\titlerunning{Generative AI for Test Driven Development: preliminary results}  % abbreviated title (for running head)
\author{Moritz Mock\orcidID{0009-0009-3156-6211} \and Jorge Melegati\orcidID{0000-0003-1303-4173} \and Barbara Russo\orcidID{0000-0003-3737-9264}}

%
%\authorrunning{Anonymous et al.} % abbreviated author list (for running head)
%\authorrunning{Mock \etal} % abbreviated author list (for running head)
%\tocauthor{Ivar Ekeland, Roger Temam, Jeffrey Dean, David Grove, Craig Chambers, Kim B. Bruce, and Elisa Bertino}
%
%TODO: uncomment for a camera-ready version 
\institute{Free University of Bozen-Bolzano, Bolzano 39100, Italy\\ \email{\{momock, jorge.melegati, brusso\}@unibz.it}}

\maketitle

\begin{abstract}
Test Driven Development (TDD) is one of the major practices of Extreme Programming for which incremental testing and refactoring trigger the code development. TDD has limited adoption in the industry, as it requires more code to be developed and experienced developers. Generative AI (GenAI) may reduce the extra effort imposed by TDD. In this work, we introduce an approach to automatize TDD by embracing GenAI either in a collaborative interaction pattern in which developers create tests and supervise the AI generation during each iteration or a fully-automated pattern in which developers only supervise the AI generation at the end of the iterations. We run an exploratory experiment with ChatGPT in which the interaction patterns are compared with the non-AI TDD regarding test and code quality and development speed. Overall, we found that, for our experiment and settings, GenAI can be efficiently used in TDD, but it requires supervision of the quality of the produced code. In some cases, it can even mislead non-expert developers and propose solutions just for the sake of the query. 
\end{abstract}

\section{Introduction}
\label{sec:introduction}
Test-driven development (TDD) is one of the major practices in Extreme Programming (XP)~\cite{Beck2022}. Nevertheless, its effectiveness is still controversial~\cite{Karac2018,Ghafari2020}. The major strength of TDD lies in its ability to deliver high-quality code through the granularity and uniformity of development~\cite{Fucci2017}. To be effective, TDD developers must have a strong command of the practice and experience in development~\cite{Causevic2011}. To facilitate its adoption, automation can be an option, and generative artificial intelligence (GenAI) tools, such as GitHub's Copilot\footnote{\url{https://copilot.microsoft.com}} and OpenAI's ChatGPT\footnote{\url{https://chat.openai.com}}, can be helpful~\cite{BavotaMenzies2022}. In particular, recent literature has shown promising results in software testing. 
For instance, Piya and Sullivan \cite{Piya2023} proposed an approach in which a test suite is fed to ChatGPT, and prompts are generated accordingly upon test failures. Liang \etal~\cite{Liang2024} have further shown that the use of GenAI can speed up testing. 
However, the quality of the generated tests and code and the role of the developers are still under discussion~\cite{Liang2024}.

In this work, we explore the use of GenAI in automating TDD and reflect on the role of the developer. 
We then perform an exploratory experiment with five developers to assess the effectiveness of our method and compare it with non-AI TDD. 

\section{Methodology}
\label{sec:methodology}

Our goal is to automatize the TDD process with GenAI, exploring which minimal knowledge is needed in each iteration and which kind of role AIs and developers may have. 
To this aim, we developed a threefold methodology: first, we defined a workflow to automate the TDD process with GenAI, then we identified interaction patterns between developers and GenAIs supporting different types of automation activities of the workflow. Finally, we implemented a tool automatizing the workflow according to the interaction patterns and then performed an experiment to compare them.
To design the workflow, we first identified the type of information that is handled in an iteration of a TDD process.
The information includes the context, the feature to be developed, the test and production code, and the execution log that was eventually output in the previous iteration. 
To obtain coherent answers, we queried ChatGPT a few times with different types of prompts. The major challenge here is to obtain an incremental output. We do not want ChatGPT to generate the code for the feature in one shot as we are not implementing Test First~\cite{Beck2022}. 
Thus, we first automated the query process by implementing a Python script that leverages OpenAI's API to use ChatGPT as GenAI. We employed the model \textit{gpt-3.5-turbo-16k}, which can have as context up to 16k tokens.
We have also explored different ways of querying the ChatGPT: (i) not mentioning the testing task at all and retrieving each message as a stand-alone, (ii) including the output of the previous query as input for the next query, or (iii) querying with all data in (i) and (ii). Based on our observation, scenario (ii), in which we send the result of the last output, is the best one. In scenario (i), ChatGPT struggled to grasp the task, and in scenario (iii), it got confused.
In the first attempts to query ChatGPT, we also observed that it had the tendency to produce the complete solution instead of performing incremental steps. To prevent from doing so, we added the sentence \textit{``stub and drivers to develop the first barely minimal test and production code''}. 
From the second iteration onwards, the additional phrase \textit{``Keep the existing tests"} at the beginning needed to be added so that the existing tests were not lost.
In the end, we were able to formulate the following prompts:
 
\begin{itemize}
   \item First iteration: Use the Assertion First pattern in TDD and stubs and drivers to develop the first barely minimal test and production code for the feature 
   $\langle feature \,\, description \rangle$ with input $\langle names\rangle$ and $\langle values\rangle$ and expected output $\langle values\rangle$
    \item Intermediate iteration: Keep the existing tests and run the next iteration of TDD to develop the barely minimal test and production code
    \item Final iteration: Refactor the code.
\end{itemize}

A second step in our methodology consists of defining the role of the developers in the TDD process automated with GenAIs.
To this aim, we defined three collaboration patterns: collaborative, fully-automated, and non-automated. 
In the \textit{collaborative pattern}, we introduce an interaction between the human developer and the AI, in which the developer is in charge of writing the test code and modifying any test or production code generated by the AI before passing it again to the AI. Then, the AI generates the production code.  
The \textit{fully-automated  pattern} automates both steps.
The developer only verifies the quality of the produced code at refactoring. 
The \textit{non-automated pattern} does not involve any AI.
Finally, we evaluated and compared the patterns with five practitioners who had experience with Python and TDD: three used the collaborative pattern, and two used the non-automated one. 
All have received the same initial exercise:

\textit{The goal of this experiment is to develop in Python the following feature: 
Develop a class TextFormatter that takes arbitrary words and horizontally center them into a line.
The class TextFormatter shall have three functions. 
The first is called setLineWidth and sets the length of the line. 
The second function receives a single word and returns the word in the center of the line. 
The third function receives two words and centers the two words in the line. 
To develop it you will use Test Driven Development and, in particular,  assertion first.}

%\end{quote} 
All participants were allowed to consult any source they liked. We recorded the screen while the participants performed their exercise. At the end of the task, they filled out a brief questionnaire.
We compared the results in terms of the number of test functions, number of assertions, test LOC, code LOC, and time to complete the task. We also inspected their test, code, and logs (in case of automation) and qualitatively evaluated the quality of what has been produced. We further collected feedback from the participants. The code produced in the experiment can be found at \url{https://github.com/moritzmock/AI4TDD}.

\section{Results}
\label{sec:results} 
\begin{figure}[t!]
    \centering
\includegraphics[trim={0 0 0 0},clip={},width=\textwidth]{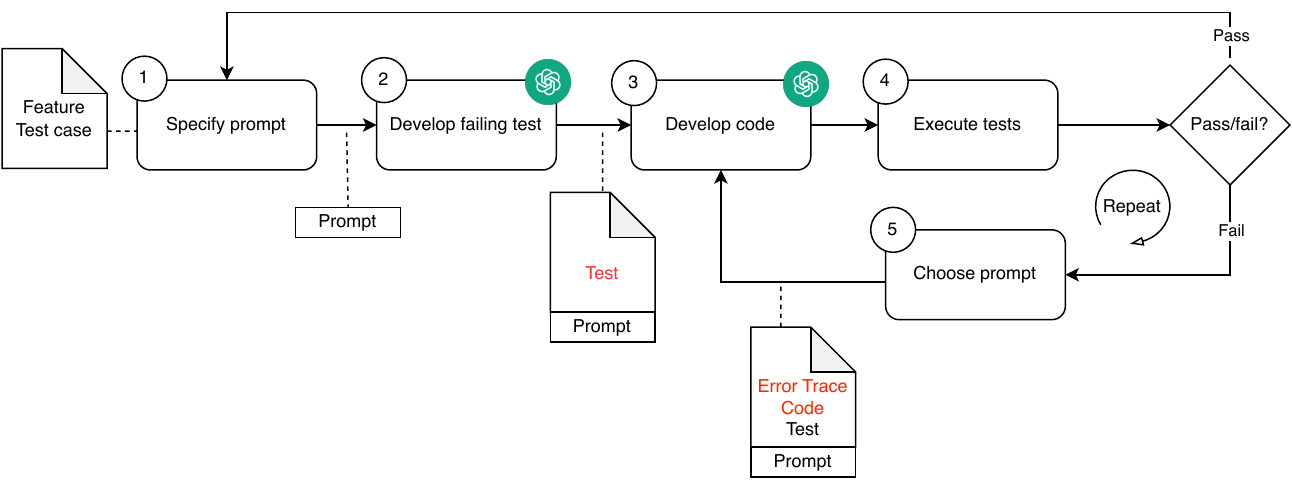}
    \caption{Fully-automated pattern}
    \label{fig:fullyAutomated}
\end{figure}

\rqone 
To answer this question, we designed two workflows, one implementing the collaborative pattern and the other fully-automated. We further implemented Python scripts that actuate them.
Fig.~\ref{fig:fullyAutomated} illustrates the workflow for the fully-automated pattern: the activities with the AI symbol are performed by the AI model, and the execution of each activity in the workflow is automated by our scripts.
The note boxes show the input needed for the next action: the prompts as described in Section~\ref{sec:methodology} and the type of data to pass to the next activity. From the response of the AI, the production code is automatically integrated, and the test suite is launched. If the AI is not able to write code that fulfils the test case(s), the prompt is resent up to five times.
In the collaborative pattern, the workflow is the same but different in its execution. Firstly, activities \circled{1} and \circled{2} are executed by the developer. Secondly, the developer can modify the input passed to the AI in activity \circled{3}. The red text in the note boxes indicates the part of the output of the previous activity that the developer can modify. 
The rest is the same as in the fully-automated pattern.

In our experiment, we  executed the two interaction patterns with five developers, whereas the authors launched the fully-automated pattern. In all cases, the experiment's task was successfully completed. The final test and production code, the process log, the screen recording, and a short feedback from the participants were collected. 

\noindent
\rqtwo~
For our exploratory experiment, we selected five developers located in two different countries with some knowledge of the programming language and experience with TDD, as shown in Table~\ref{tab:demographic}. 

\begin{table}[b]
    \centering
    \renewcommand{\arraystretch}{1.2}
    \setlength{\tabcolsep}{4pt}
    \footnotesize
    \caption{Demographics of the developers}
    \label{tab:demographic}
    \begin{tabular*}{\textwidth}{@{\extracolsep{\fill}}lcccc@{}}
        \hline
         & \makecell{TDD \\ Experience} & \makecell{Python \\ Experience}& Role & \makecell{Interaction\\Scenario} \\ \hline
         P1 & \makecell[c]{$>$ 3 years} & $<$ 1 year & \makecell[c]{Software developer} & collaborative\\
         P2 & \makecell[c]{1-3 years} & \makecell[c]{$>$ 3 years} & \makecell[c]{Data scientist} & collaborative \\
         P3 & \makecell[c]{1-3 years} & \makecell[c]{$>$ 3 years} & \makecell[c]{Software developer} & non-automated \\
          P4&$<$ 1 year&$<$ 1 years&Software engineering student&collaborative\\
         P5&1-3 years&$<$ 1 years&Software developer&non-automated\\
         \hline
    \end{tabular*}
\end{table}

Three followed the collaborative pattern, and two followed the non-automated one. We compared the results of the experiment in terms of the metrics defined in Section~\ref{sec:methodology}. Table~\ref{tab:measuresExperiments} reports the results for the five participants P1-P5 and the fully-automated pattern, F1.
For F1, the authors launched the task a few times. The first times were used to learn the type of prompts needed to automate TDD iteratively, and the last one was executed to compare the results of the fully-automated with the ones of the other patterns. Table~\ref{tab:measuresExperiments} reports the value for the last run. It is worth noticing that the AI acts as a tester and developer, so we are able to log the activities of both. In some iterations, the developer implemented the code with no interaction with a tester, as may happen in real cases. 

\begin{table}[!t]
    \centering
    \footnotesize
    \caption{Evaluation of the results for the participants P1, P2, and P3 and the fully-automated one F1. $\#$ of LOC does not contain blank lines.}
\label{tab:measuresExperiments}
    \begin{tabular*}{\textwidth}{@{\extracolsep{\fill}}lcccccc@{}}
        \hline
         & \makecell{\# \\test functions} & \makecell{\# \\assertions} & \makecell{test LOC} & \makecell{code LOC} & \makecell{Time  to\\complete }&\makecell{$\#$ \\ iterations}\\\hline
        P1 & 7 & 7 & 34 & 25 & 30 min.&32\\
        P2 & 3 & 9 & 19 & 19 & 30 min.&49\\
        P3 & 11 & 17 & 69 & 31 & 35 min.&NA\\
        P4&4&4&16&12&40 min.&44\\
        P5&3&3&16&14&40 min.&NA\\
        F1 & 1 & 3 & 14 & 17 & 12 min. & 8\\\hline
    \end{tabular*}
\end{table}

\par\noindent\textbf{Test coverage and code inspection.}
All participants managed to complete the tasks within the time limit of 40 minutes. They created different sets of tests and considered different edge cases. 
P1 did not develop tests for all the valid inputs but implemented the edge cases where words are larger than the line width or empty and a new functionality that cuts the word if its length is larger than the line. P2 did not create a class but three individual functions instead. In this case, our script was not able to parse the messages received from ChatGPT and extract the generated code. Thus, P2 had to struggle a bit until he was able to understand the automation. The final code was clean, with only assertions per test function, but it contained one redundant assertion per function and no edge cases.
P3 implemented the largest number of test cases, including all valid and invalid cases and specific exceptions: \textit{InvalidLineWidthException}, \textit{CenteringNotPossibleError}, and \textit{SpreadNotPossibleError}.
The final code is not completely clean, though, as it includes repeated assertions for one method and more than one test method for the same function to test. 
P4 implemented all test cases for all functions but no edge cases, and the final code is not correct (see Log inspection, next paragraph).
P5 worked with no assistance from the AI and implemented only one test case per function, with valid input and no edge cases.
F1 produced one test function with three assertions, each testing a valid input. One valid case was missing and no edge case or invalid value was tested. F1 was the fastest to complete the task.
\par\noindent\textbf{Log and screen inspection.} For all developers who collaborated with the AI, we logged the interactions and watched the screens' recordings. The last column of Table~\ref{tab:measuresExperiments} reports the number of interactions. At each interaction, we logged the test and production code, the execution trace, and the changes made by the developer to fix the AI output and make the test pass. All the changes have finally produced a correct code except in the case of P4.
ChatGPT recommended these changes at the final interaction:
%\vspace{-5pt}
\begin{lstlisting}[
language = Python,
numbers=none,
xleftmargin=0ex]
Changes made to the code:
1. In the `test_TwoWordSpreadEven` test, the `TextFormatter` variable was mistakenly assigned instead of `textFormatter`.
2. In the `test_TwoWordSpreadOdd` test, the `TextFormatter` variable was mistakenly assigned instead of `textFormatter`.
3. In the `test_WordCenterOdd` test, the expected result was corrected by removing the extra spaces. 
\end{lstlisting}
However, the proposed changes do not fix the bug but simply avoid the execution of the code revealing the bug. 
For F1, no test failed, suggesting that the AI was more concerned with not failing the test than with developing a high-quality solution. 

\par\noindent\textbf{Analysing feedback.} We finally asked the participants about their experience with AI. The answers are reported in Table~\ref{tab:feedback}.
\begin{table}[!b]
    \centering
    \footnotesize
        \caption{Feedback}
    \label{tab:feedback}
    \begin{tabular}{clp{0.85\textwidth}}
    \hline
        & \makecell{Perceived\\ difficulty} & Feedback  \\\hline
        P1&Easy  & To be honest, the ``presence" of the AI made me a little unsure in the beginning, because I was concerned about its behavior and if I should adapt to fit its need. Once I realized the AI would adapt to my needs (in particular my dev-flow), I think the experience went way more smoothly.\\
        P2&Easy &The tool did not work as expected. It seemed kinda buggy, as it did not add any code to the existing file. I was expecting more from an AI tool as normally ChatGPT is able to complete such trivial tasks. \\
        P3&Easy &It was fun, the requirements are very broad, so maybe the assumptions can vary a bit from person to person.\\
        P4&Fine &NA\\
        P5&Hard &NA\\
        \hline
    \end{tabular}
\end{table}
The positive feeling described by P1 is related to the compatibility of the tool with the way P1 works. This feeling of comfort is known to be a key determinant for acceptance and adoption of new technologies and methodologies in software engineering~\cite{Riemenschneider2002}. This aspect will also be important for the adoption of AI-based tools and, as such, should be considered in the development of this new generation of tools.
P2, who created a set of functions rather than a class, got frustrated when the GenAI tool did not work as expected. Apparently, the same frustration has also been observed in students in a study on using GenAI tools for teaching software engineering~\cite{Choudhuri2024}.
In this case, though, it was not the GenAI that did not process the query as expected but rather our script, which was not designed to extract functions from the messages of ChatGPT. Of course, the developer could not distinguish the difference.  
Clearly, P2 did not read the task's description, which required developing a class, but this gave us the hint to refactor our tool so that it is now also capable of extracting functions from ChatGPT messages.  
Also P3 had a positive experience in performing the task autonomously and even suggested the authors some refinement of the experiment. Being expert in both TDD and Python the task was not hard, and the resulting code was the most creative. 
No feedback was obtained from P4 and P5.
Overall, we found that ChatGPT can meet the expectations of the developers in assisting in their job, but without replacing developers in terms of creativity and quality of the code. To obtain satisfactory collaboration with developers, AI should be well integrated into the automation of development activities. The solutions generated may be incomplete or buggy, and non-expert developers may not notice this and trust the AI straight away. 

\section{Related work}
\label{sec:relatedWork}
In this section, we present research works on GenAI for software testing and TDD, in particular. 
Bird \etal~\cite{Bird2022} analyzed forum discussions from early GitHub's Copilot users, collected their impressions on the tool usage, and observed that support in writing unit tests was one of the major benefits. Producing test cases quickly was the major result of a large survey on the usability of AI programming assistants of Liang \etal~\cite{Liang2024}. They reported that finding edge cases in testing was among the major reasons for using AI. On the other hand, verifying AI answers (e.g., to meet software requirements) was the major reason for not using them. Guilherme and Vincenzi~\cite{VitorAuri2023} used  OpenAI API to generate unit tests and concluded that the tool has a good performance in terms of mutation score and code coverage. GPTDroid~\cite{Liu2024} uses ChatGPT for GUI testing of mobile apps as a Q\&A task and was able to achieve higher coverage and greater efficiency in finding bugs. 
Lahiri \etal~\cite{Lahiri2023} propose ITDCG, a workflow with Open AI's Codex for interactive test-driven code generation. 
Tests and code are generated simultaneously, not incrementally and iteratively, as in TDD. No particular mention is made of how to query the AI or the role of the developer. 
Tian and Chen~\cite{Tian2023} introduce Test-case-driven Chain of Thought (TCoT), an approach for improving code generation by using the description of the tests in natural language. 
The results are promising; however, they did not focus on TDD or any iterative testing process. 
Piya and Sullivan~\cite{Piya2023} introduced the LLM4TDD framework to incorporate GenAI into TDD. A developer develops within a coding environment that interacts with a GenAI. The developer manually copies, if needed, code and tests and whether the latter ones fail. 
An evaluation of the framework reached a success rate of 88.5\%.
The authors also identified best practices to ensure that ChatGPT solves the correct problem and to reduce the effort. Our work follows these recommendations but also provides 1) different ways of interactions between humans and AI, 2) a structure for the input with predefined prompts to avoid generating unwanted code, 3) a control layer for the collaborative pattern in which the quality of the code is iteratively verified by the developer. The experiment with real developers helped to understand the issue of such collaboration at a fine granularity level.
\section{Conclusions}
\label{sec:conclusions}
In this work, we defined interaction patterns between developers and GenAIs for the automation of TDD. 
We conducted an exploratory experiment with practitioners to evaluate the feasibility of our automation and the quality of the produced solutions. Overall, we found that for our experiment and settings, GenAI can be efficiently used in TDD, but it requires supervision on the quality of the code produced. In some cases, it can mislead non-expert developers and propose solutions that change tests rather than the production code, which may remain buggy, to make tests pass. 
In future work, we will extend our methodology to incorporate other interaction patterns (e.g., the developer can choose freely how to query the GenAI), different automation and GenAI, and involve a larger number of practitioners both in the experiment and in the feedback.

\section*{Acknowledgments}
\label{sec:ack}
We thank the practitioners who participated in the study for their valuable contribution. 
Moritz Mock is partially funded by the National Recovery and Resilience Plan (Piano Nazionale di Ripresa e Resilienza, PNRR - DM 117/2023).

\bibliographystyle{splncs04}
\bibliography{refs}

\end{document}